\documentclass[hyper,a4paper,12pt,oneside]{JHEP3}
\title{Density Matrix Deformations in Quantum and
Statistical Mechanics at Planck-Scale}
\author{ A.E.Shalyt-Margolin \\   National
Center of Particles and High Energy Physics, Bogdanovich Str. 153,
Minsk 220040,
 Belarus \\ E-mail: \email{a.shalyt@mail.ru, alexm@hep.by}}
 \abstract{In this work the Quantum and Statistical Mechanics of the Early
Universe, i.e. at Planck scale, is considered as a deformation of
the well-known theories. In so doing the primary object under
deformation in both cases is the density matrix. It is
demonstrated that in construction of the deformed quantum
mechanical and statistical density matrices referred to as density
pro-matrices there is a complete analog. The principal difference
lies in a nature of the deformation parameter that is associated
with the fundamental length in the first case and with a maximum
temperature in the second case. Consideration is also given to
some direct consequences, specifically the use of the explicitly
specified exponential ansatz in the derivation from the basic
principles of a semiclassical Bekenstein-Hawking formula for the
black hole entropy and of the high-temperature complement to the
canonical Gibbs distribution }

\begin{document}
\section{Introduction}
As is known, at Planck scale Quantum Mechanics undergoes
variation: it should be subjected to deformation. This is realized
due to the   presence of the Generalized Uncertainty Relations
(GUR) and hence the fundamental length \cite{r1},\cite{r2}. The
deformation in Quantum Mechanics at Planck scale takes different
paths: commutator deformation \cite{r3},
\cite{r4},\cite{r5},\cite{r6} or density matrix deformation
\cite{r7}, \cite{r8}. In the present work the second approach is
extended by the author to the Statistical Mechanics at Plank
scale.And similar to the quantum mechanics, the primary object is
the density matrix. A complete analog in construction of the
deformed quantum mechanical and statistical density matrices is
revealed. The principal difference is the deformation parameter
that is associated with the fundamental length in  the first case
and with a maximum temperature in the second case. Consideration
is given to some consequences including the use of the explicitly
specified exponential ansatz in the derivation from the basic
principles of a semiclassical Bekenstein-Hawking formula for the
black hole entropy and a high- temperature complement to the
canonical Gibbs distribution. In conclusion possible applications
of the obtained results are discussed.

\section {Density Matrix Deformation in Quantum Mechanics  at Planck
Scale}

In this section the principal features of QMFL construction with the
 use of the density matrix deformation are briefly outlined \cite{r7},
 \cite{r8}.
  As mentioned above, for the fundamental deformation parameter
we use $\alpha = l_{min}^{2 }/x^{2 }$ where $x$ is the scale. In
contrast with \cite{r7}, \cite{r8}, for the deformation parameter
we use $\alpha$ rather than  $\beta$ to avoid confusion, since
quite a distinct value is denoted by $\beta$ in Statistical
Mechanics:$\beta=1/k_{B}T$.
\\
\noindent {\bf Definition 1.} {\bf(Quantum Mechanics with
Fundamental Length)}
\\
\noindent Any system in QMFL is described by a density pro-matrix
of the form $${\bf
\rho(\alpha)=\sum_{i}\omega_{i}(\alpha)|i><i|},$$ where
\begin{enumerate}
\item $0<\alpha\leq1/4$;
\item The vectors $|i>$ form a full orthonormal system;
\item $\omega_{i}(\alpha)\geq 0$ and for all $i$  the
finite limit $\lim\limits_{\alpha\rightarrow
0}\omega_{i}(\alpha)=\omega_{i}$ exists;
\item
$Sp[\rho(\alpha)]=\sum_{i}\omega_{i}(\alpha)<1$,
$\sum_{i}\omega_{i}=1$;
\item For every operator $B$ and any $\alpha$ there is a
mean operator $B$ depending on  $\alpha$:\\
$$<B>_{\alpha}=\sum_{i}\omega_{i}(\alpha)<i|B|i>.$$

\end{enumerate}
Finally, in order that our definition 1 agree with the result of
section 2, the following condition must be fulfilled:
\begin{equation}\label{U1}
Sp[\rho(\alpha)]-Sp^{2}[\rho(\alpha)]\approx\alpha.
\end{equation}
Hence we can find the value for $Sp[\rho(\alpha)]$ satisfying the
condition of definition 1:
\begin{equation}\label{U2}
Sp[\rho(\alpha)]\approx\frac{1}{2}+\sqrt{\frac{1}{4}-\alpha}.
\end{equation}

According to point 5),  $<1>_{\alpha}=Sp[\rho(\alpha)]$. Therefore
for any scalar quantity $f$ we have $<f>_{\alpha}=f
Sp[\rho(\alpha)]$. In particular, the mean value
$<[x_{\mu},p_{\nu}]>_{\alpha}$ is equal to
\\
$$<[x_{\mu},p_{\nu}]>_{\alpha}= i\hbar\delta_{\mu,\nu}
Sp[\rho(\alpha)]$$
\\
We denote the limit $\lim\limits_{\alpha\rightarrow
0}\rho(\alpha)=\rho$ as the density matrix. Evidently, in the
limit $\alpha\rightarrow 0$ we return to QM.

As follows from definition 1,
$<(j><j)>_{\alpha}=\omega_{j}(\alpha)$, from whence the
completeness condition by $\alpha$ is
\\$<(\sum_{i}|i><i|)>_{\alpha}=<1>_{\alpha}=Sp[\rho(\alpha)]$. The
norm of any vector $|\psi>$ assigned to  $\alpha$ can be defined
as
\\$<\psi|\psi>_{\alpha}=<\psi|(\sum_{i}|i><i|)_{\alpha}|\psi>
=<\psi|(1)_{\alpha}|\psi>=<\psi|\psi> Sp[\rho(\alpha)]$, where
$<\psi|\psi>$ is the norm in QM, i.e. for $\alpha\rightarrow 0$.
Similarly, the described theory may be interpreted using a
probabilistic approach, however requiring  replacement of $\rho$
by $\rho(\alpha)$ in all formulae.

\renewcommand{\theenumi}{\Roman{enumi}}
\renewcommand{\labelenumi}{\theenumi.}
\renewcommand{\labelenumii}{\theenumii.}

It should be noted:

\begin{enumerate}
\item The above limit covers both Quantum
and Classical Mechanics. Indeed, since $\alpha\sim L_{p}^{2 }/x^{2
}=G \hbar/c^3 x^{2}$, we obtain:
\begin{enumerate}
\item $(\hbar \neq 0,x\rightarrow
\infty)\Rightarrow(\alpha\rightarrow 0)$ for QM;
\item $(\hbar\rightarrow 0,x\rightarrow
\infty)\Rightarrow(\alpha\rightarrow 0)$ for Classical Mechanics;
\end{enumerate}
\item As a matter of fact, the deformation parameter $\alpha$
should assume the value $0<\alpha\leq1$.  However, as seen from
(\ref{U2}), $Sp[\rho(\alpha)]$ is well defined only for
$0<\alpha\leq1/4$, i.e. for $x=il_{min}$ and $i\geq 2$ we have no
problems at all. At the point, where $x=l_{min}$, there is a
singularity related to complex values assumed by
$Sp[\rho(\alpha)]$ , i.e. to the impossibility of obtaining a
diagonalized density pro-matrix at this point over the field of
real numbers. For this reason definition 1 has no sense at the
point $x=l_{min}$.

\item We consider possible solutions for (\ref{U1}).
For instance, one of the solutions of (\ref{U1}), at least to the
first order in $\alpha$, is $$\rho^{*}(\alpha)=\sum_{i}\alpha_{i}
exp(-\alpha)|i><i|,$$ where all $\alpha_{i}>0$ are independent of
$\alpha$  and their sum is equal to 1. In this way
$Sp[\rho^{*}(\alpha)]=exp(-\alpha)$. Indeed, we can easily verify
that \begin{equation}\label{U3}
Sp[\rho^{*}(\alpha)]-Sp^{2}[\rho^{*}(\alpha)]=\alpha+O(\alpha^{2}).
\end{equation}
 Note that in the momentum representation $\alpha=p^{2}/p^{2}_{pl}$,
where $p_{pl}$ is the Planck momentum. When present in matrix
elements, $exp(-\alpha)$ can damp the contribution of great
momenta in a perturbation theory.
\item It is clear that within the proposed description the
states with a unit probability, i.e. pure states, can appear only
in the limit $\alpha\rightarrow 0$, when all $\omega_{i}(\alpha)$
except for one are equal to zero or when they tend to zero at this
limit. In our treatment pure state are states, which can be
represented in the form $|\psi><\psi|$, where $<\psi|\psi>=1$.

\item We suppose that all the definitions concerning a
density matrix can be transferred to the above-mentioned
deformation of Quantum Mechanics (QMFL) through changing the
density matrix $\rho$ by the density pro-matrix $\rho(\alpha)$ and
subsequent passage to the low energy limit $\alpha\rightarrow 0$.
Specifically, for statistical entropy we have
\begin{equation}\label{U4}
S_{\alpha}=-Sp[\rho(\alpha)\ln(\rho(\alpha))].
\end{equation}
The quantity of $S_{\alpha}$ seems never to be equal to zero as
$\ln(\rho(\alpha))\neq 0$ and hence $S_{\alpha}$ may be equal to
zero at the limit $\alpha\rightarrow 0$ only.
\end{enumerate}
Some Implications:
\begin{enumerate}
\item If we carry out measurement on the pre-determined scale, it is
impossible to regard the density pro-matrix as a density matrix
with an accuracy better than particular limit $\sim10^{-66+2n}$,
where $10^{-n}$ is the measuring scale. In the majority of known
cases this is sufficient to consider the density pro-matrix as a
density matrix. But on Planck's scale, where the quantum
gravitational effects and Plank energy levels cannot be neglected,
the difference between $\rho(\alpha)$ and  $\rho$ should be taken
into consideration.

\item Proceeding from the above, on Planck's scale the
notion of Wave Function of the Universe (as introduced in
\cite{r9}) has no sense, and quantum gravitation effects in this
case should be described with the help of density pro-matrix
$\rho(\alpha)$ only.
\item Since density pro-matrix $\rho(\alpha)$ depends on the measuring
scale, evolution of the Universe within the inflation model
paradigm \cite{r10} is not a unitary process, or otherwise the
probabilities $p_{i}=\omega_{i}(\alpha)$  would be preserved.
\end{enumerate}

\section {Deformed Density Matrix in Statistical Mechanics at Planck
Scale}

To begin, we recall the generalized uncertainty relations
 "coordinate - momentum" \cite{r4},\cite{r5},\cite{r6}:
\begin{equation}\label{U5}
\triangle x\geq\frac{\hbar}{\triangle p}+\alpha^{\prime}
L_{p}^2\frac{\triangle p}{\hbar}.
\end{equation}
 Using relations (\ref{U5}) it is easy to obtain a similar relation for the
 "energy - time" pair. Indeed (\ref{U5}) gives
\begin{equation}\label{U6}
\frac{\Delta x}{c}\geq\frac{\hbar}{\Delta p c
}+\alpha^{\prime}
L_{p}^2\,\frac{\Delta p}{c \hbar},
\end{equation}
then
\begin{equation}\label{U7}
\Delta t\geq\frac{\hbar}{\Delta
E}+\alpha^{\prime}\frac{L_{p}^2}{c^2}\,\frac{\Delta p
c}{\hbar}=\frac{\hbar}{\Delta
E}+\alpha^{\prime}
t_{p}^2\,\frac{\Delta E}{ \hbar}.
\end{equation}
where the smallness of $L_p$ is taken into account so that the
difference between $\Delta E$ and $\Delta (pc)$ can be neglected
and $t_{p}$  is the Planck time
$t_{p}=L_p/c=\sqrt{G\hbar/c^5}\simeq 0,54\;10^{-43}sec$. From
whence it follows that we have a  maximum energy of the order of
Planck's:
\\
$$E_{max}\sim E_{p}$$
\\
Proceeding to the Statistical Mechanics, we further assume that an
internal energy of any ensemble U could not be in excess of
$E_{max}$ and hence temperature $T$ could not be in excess of
$T_{max}=E_{max}/k_{B}\sim T_{p}$. Let us consider density matrix
in Statistical Mechanics (see \cite{r11}, Section 2, Paragraph 3):
\begin{equation}\label{U8}
\rho_{stat}=\sum_{n}\omega_{n}|\varphi_{n}><\varphi_{n}|,
\end{equation}
where the probabilities are given by
\\
$$\omega_{n}=\frac{1}{Q}\exp(-\beta E_{n})$$ and
\\
$$Q=\sum_{n}\exp(-\beta E_{n})$$
\\
Then for a canonical Gibbs ensemble the value
\begin{equation}\label{U9}
\overline{\Delta(1/T)^{2}}=Sp[\rho_{stat}(\frac{1}{T})^{2}]
-Sp^{2}[\rho_{stat}(\frac{1}{T})],
\end{equation}
is always equal to zero, and this follows from the fact that
$Sp[\rho_{stat}]=1$. However, for very high temperatures $T\gg0$
we have $\Delta (1/T)^{2}\approx 1/T^{2}\geq 1/T_{max}^{2}$. Thus,
for $T\gg0$ a statistical density matrix $\rho_{stat}$ should be
deformed so that in the general case
\begin{equation}\label{U10}
Sp[\rho_{stat}(\frac{1}{T})^{2}]-Sp^{2}[\rho_{stat}(\frac{1}{T})]
\approx \frac{1}{T_{max}^{2}},
\end{equation}
or \begin{equation}\label{U11}
Sp[\rho_{stat}]-Sp^{2}[\rho_{stat}] \approx
\frac{T^{2}}{T_{max}^{2}},
\end{equation}
In this way $\rho_{stat}$ at very high $T\gg 0$ becomes dependent
on the parameter $\tau = T^{2}/T_{max}^{2}$, i.e. in the most
general case
\\
$$\rho_{stat}=\rho_{stat}(\tau)$$ and $$Sp[\rho_{stat}(\tau)]<1$$
\\
and for $\tau\ll 1$ we have $\rho_{stat}(\tau)\approx\rho_{stat}$
(formula (\ref{U8})) .\\ This situation is identical to the case
associated with the deformation parameter $\alpha = l_{min}^{2
}/x^{2}$ of QMFL given in section â 2 \cite{r8}. That is the
condition $Sp[\rho_{stat}(\tau)]<1$ has an apparent physical
meaning when:
\begin{enumerate}
 \item At temperatures close to $T_{max}$ some portion of information
about the ensemble is inaccessible in accordance with the
probability that is less than unity, i.e. incomplete probability.
 \item And vice versa, the longer is the distance from $T_{max}$ (i.e.
when approximating the usual temperatures), the greater is the
bulk of information and the closer is the complete probability to
unity.
\end{enumerate}
 Therefore similar to the introduction of the deformed
quantum-mechanics density matrix in section 3 \cite{r8} and
previous section of this paper,we give the following
\\
\noindent {\bf Definition 2.} {\bf(Deformation of Statistical
Mechanics)} \noindent \\Deformation of Gibbs distribution valid
for temperatures on the order of the Planck's $T_{p}$ is described
 by deformation of a statistical density matrix
  (statistical density pro-matrix) of the form
\\$${\bf \rho_{stat}(\tau)=\sum_{n}\omega_{n}(\tau)|\varphi_{n}><\varphi_{n}|}$$
 having the deformation parameter
$\tau = T^{2}/T_{max}^{2}$, where
\begin{enumerate}
\item $0<\tau \leq 1/4$;
\item The vectors $|\varphi_{n}>$ form a full orthonormal system;
\item $\omega_{n}(\tau)\geq 0$ and for all $n$ at $\tau \ll 1$
 we obtain
 $\omega_{n}(\tau)\approx \omega_{n}=\frac{1}{Q}\exp(-\beta E_{n})$
In particular, $\lim\limits_{T_{max}\rightarrow \infty
(\tau\rightarrow 0)}\omega_{n}(\tau)=\omega_{n}$
\item
$Sp[\rho_{stat}(\tau)]=\sum_{n}\omega_{n}(\tau)<1$,
$\sum_{n}\omega_{n}=1$;
\item For every operator $B$ and any $\tau$ there is a
mean operator $B$ depending on  $\tau$ \\
$$<B>_{\tau}=\sum_{n}\omega_{n}(\tau)<n|B|n>.$$
\end{enumerate}
Finally, in order that our Definition 2 agree with the formula
(\ref{U11}), the following condition must be fulfilled:
\begin{equation}\label{U12}
Sp[\rho_{stat}(\tau)]-Sp^{2}[\rho_{stat}(\tau)]\approx \tau.
\end{equation}
Hence we can find the value for $Sp[\rho_{stat}(\tau)]$
 satisfying
the condition of Definition 2 (similar to Definition 1):
\begin{equation}\label{U13}
Sp[\rho_{stat}(\tau)]\approx\frac{1}{2}+\sqrt{\frac{1}{4}-\tau}.
\end{equation}
It should be noted:

\begin{enumerate}
\item The condition $\tau \ll 1$ means that $T\ll T_{max}$ either
$T_{max}=\infty$ or both in accordance with a normal Statistical
Mechanics and canonical Gibbs distribution (\ref{U8})
\item Similar to QMFL in Definition 1, where the deformation
parameter $\alpha$ should assume the value $0<\alpha\leq1/4$. As
seen from (\ref{U13}), here $Sp[\rho_{stat}(\tau)]$ is well
defined only for $0<\tau\leq1/4$. This means that the feature
occurring in QMFL at the point of the fundamental length
$x=l_{min}$ in the case under consideration is associated with the
fact that {\bf highest  measurable temperature of the ensemble is
always} ${\bf T\leq \frac{1}{2}T_{max}}$.

\item The constructed deformation contains all four fundamental constants:
 $G,\hbar,c,k_{B}$ as $T_{max}=\varsigma T_{p}$,where $\varsigma$
 is the denumerable function of  $\alpha^{\prime}$
(\ref{U5})and $T_{p}$, in its turn, contains all the above-mentioned
 constants.

\item Again similar to QMFL, as a possible solution for (\ref{U12})
we have an exponential ansatz
\\
$$\rho_{stat}^{*}(\tau)=\sum_{n}\omega_{n}(\tau)|n><n|=\sum_{n}
exp(-\tau) \omega_{n}|n><n|$$
\\
\begin{equation}\label{U14}
Sp[\rho_{stat}^{*}(\tau)]-Sp^{2}[\rho_{stat}^{*}(\tau)]=\tau+O(\tau^{2}).
\end{equation}
In such a way with the use of an exponential ansatz (\ref{U14})
the deformation of a canonical Gibbs distribution at Planck scale
(up to factor $1/Q$) takes an elegant and completed form:
\begin{equation}\label{U15}
{\bf \omega_{n}(\tau)=exp(-\tau)\omega_{n}= exp(-\frac{T^{2}}
{T_{max}^{2}}-\beta E_{n})}
\end{equation}
where $T_{max}= \varsigma T_{p}$

\section{Comments to the Bekenstein-Hawking Formula and Measuring
Procedure}
 It should be noted that an exponential ansatz yielding a result
in case of the statistical mechanics, in quantum mechanics may be
used in the derivation of Bekenstein-Hawking formula for the black
hole entropy in semiclassical approximation from the basic
principles \cite{r8}:
 In the process factor 1/4 is interpreted as a growing density of the
entropy for the observer at the conventional scales when measuring
is performed close to the singularity. Also note that (\ref{U3})
may be written as a series
 \begin{equation}\label{U15}
Sp[\rho(\alpha)]-Sp^{2}[\rho(\alpha)]=\alpha+a_{0}\alpha^{2}
+a_{1}\alpha^{3}+...
\end{equation}
As a result, a measurement procedure using the exponential ansatz
may be understood as the calculation of factors
$a_{0}$,$a_{1}$,... or the definition of additional members in the
exponent which "destroy" $a_{0}$,$a_{1}$,... . It is easy to check
that the exponential ansatz gives $a_{0}=-3/2$, being coincident
with the logarithmic correction factor for the black hole entropy
\cite{r15}.

\section{Conclusion}

Thus, it has been shown that between the deformations of quantum
and statistical mechanics of the early Universe there is a
complete analog. Some consequences have been demonstrated. Of
interest are possible applications presented in the work. Of
particular interest is also the problem of a rigorous proof for
the generalized uncertainty relations (GUR) in thermodynamics
\cite{r12},\cite{r13} as a complete analog of the corresponding
relations in Quantum Mechanics \cite{r1}, \cite{r3,r4,r5,r6}. The
methods developed in this study may be interesting for the
investigation of black hole thermodynamics \cite{r14} based on GUR
(\ref{U5}).These problems will be considered by the author in
subsequent papers.

\end{enumerate}

%References

\end{document}